\documentclass[a4paper,12pt]{article}

\usepackage{amsmath,amssymb,amsfonts,epsfig,cite}
\usepackage[paper=letterpaper,margin=1.2in]{geometry}

\begin{document}


\begin{flushright}
Imperial/TP/09/AH/02 \\
NSF-KITP-09-40 \\
\end{flushright}
\vskip 0.25in

\renewcommand{\thefootnote}{\fnsymbol{footnote}}
\centerline{\Huge Chern-Simons: Fano and Calabi-Yau}

\vskip 2mm
\centerline{
Amihay Hanany${}^{1}$\footnote{\tt a.hanany@imperial.ac.uk}
and
Yang-Hui He${}^{2}$\footnote{\tt hey@maths.ox.ac.uk}
}
~\\
~\\
{\hspace{-1in}
\scriptsize
\begin{tabular}{ll}
  ${}^1$ 
  &{\it Theoretical Physics Group, Blackett Laboratory, 
    Imperial College, London SW7 2AZ, U.K.;}\\
  &{\it KITP, University of California, 
    Santa Barbara, CA 93106-4030, USA}\\
  ${}^2$
  & {\it Rudolf Peierls Centre for Theoretical Physics, Oxford University, 1 Keble Road, OX1 3NP, U.K.;}\\
  & {\it Collegium Mertonense in Academia Oxoniensi, Oxford, OX1 4JD, U.K.;}\\
  & {\it Mathematical Institute, Oxford University, 24-29 St.\ Giles', Oxford, OX1 3LB, U.K.}\\
\end{tabular}
}

\newcommand{\todo}[1]{{\bf ?????!!!! #1 ?????!!!!}\marginpar{$\Longleftarrow$}}
\newcommand{\beq}{\begin{equation}}
\newcommand{\eeq}{\end{equation}}
\newcommand{\ba}{\begin{array}}
\newcommand{\ea}{\end{array}}
\newcommand{\bea}{\begin{eqnarray}}
\newcommand{\eea}{\end{eqnarray}}
\newcommand{\bean}{\begin{eqnarray*}}
\newcommand{\eean}{\end{eqnarray*}}
\newcommand{\eref}[1]{(\ref{#1})}
\newcommand{\sref}[1]{\S\ref{#1}}
\newcommand{\fref}[1]{Figure~\ref{#1}}
\newcommand{\nn}{\nonumber}
\newcommand{\tr}{\mathop{\rm Tr}}
\newcommand{\comment}[1]{}

\newcommand{\CM}{{\cal M}}
\newcommand{\CN}{{\cal N}}
\newcommand{\CZ}{{\cal Z}}
\newcommand{\cO}{{\cal O}}
\newcommand{\cB}{{\cal B}}
\newcommand{\cC}{{\cal C}}
\newcommand{\cD}{{\cal D}}
\newcommand{\cE}{{\cal E}}
\newcommand{\cF}{{\cal F}}
\newcommand{\cX}{{\cal X}}
\newcommand{\IP}{\mathbb{P}}
\newcommand{\IR}{\mathbb{R}}
\newcommand{\IC}{\mathbb{C}}
\newcommand{\IF}{\mathbb{F}}
\newcommand{\IV}{\mathbb{V}}
\newcommand{\IZ}{\mathbb{Z}}
\newcommand{\f}{{\cal F}^{\flat}}
\newcommand{\firr}[1]{{}^{{\rm Irr}}\!{\cal F}^{\flat}_{#1}}

\newcommand{\vev}[1]{\langle #1 \rangle}
\newtheorem{theorem}{\bf THEOREM}
\def\thetheorem{\thesection.\arabic{theorem}}
\newtheorem{conjecture}{\bf CONJECTURE}
\def\thetheorem{\thesection.\arabic{conjecture}}
\newtheorem{observation}{\bf OBSERVATION}
\def\thetheorem{\thesection.\arabic{observation}}

\def\theequation{\thesection.\arabic{equation}}
\newcommand{\setall}{\setcounter{equation}{0}
        \setcounter{theorem}{0}}
\newcommand{\setequation}{\setcounter{equation}{0}}

\newcommand{\bull}{$\bullet\ $}
\newcommand{\CP}{\mathbb P}
\newcommand{\tmat}[1]{{\tiny \left(\begin{matrix} #1 \end{matrix}\right)}}

\begin{abstract}
We present the complete classification of smooth toric Fano threefolds, known to the algebraic geometry literature, and perform some preliminary analyses in the context of brane-tilings and Chern-Simons theory on M2-branes probing Calabi-Yau fourfold singularities. We emphasise that these 18 spaces should be as intensely studied as their well-known counter-parts: the del Pezzo surfaces.
\end{abstract}

\newpage
\tableofcontents

\section{Introduction}
A flurry of activity has, since the initial work of Bagger-Lambert \cite{BaggerLambert} and of Gustavsson \cite{Gustavsson:2007vu}, rather excited the community for the past two years upon the subject of supersymmetric Chern-Simons theories. It is by now widely believed that the world-volume theory of M2-branes on various back-grounds is given by a $(2+1)$-dimensional quiver Chern-Simons (QCS) theory \cite{rams,Mukhi:2008ux,Aharony:2008ug,Benna:2008zy,Kim:2007ic,Hosomichi:2008jb,Schnabl,Martelli:2008si,Hanany:2008cd,Hanany:2008fj,Imamura:2008nn,Ueda:2008hx,Franco:2008um,Hanany:2008gx,Aharony:2008gk,Franco,Ami,Krishnan:2009nw,Amariti:2009rb,Imamura:2009ph}, most conveniently described by a brane tiling.

Even though analogies with the case of D3-branes in Type IIB, whose world-volume theory is a $(3+1)$-dimensional supersymmetric quiver gauge theory, are very re-assuring, the story is much less understood for the M2 case. Much work has been devoted to the understanding of issues such as orbifolding, phases of duality, brane-tilings and dimer/crystal models etc. Nevertheless, the r\^ole played by the correspondence between the world-volume theory and the underlying Calabi-Yau geometry is of indubitable importance. Indeed, there is a bijection: the vacuum moduli space of the former is, tautologcally, the latter, while the geometrical engineering on the latter gives, by construction, the former. This bijection, called respectively the ``forward'' and ``inverse'' algorithms \cite{Feng:2000mi}, persists in any dimension and can be succintly summarised in Table \ref{t:embed}.

\begin{table}[h!]
\begin{center}
\begin{tabular}{|c|c|c|c|c|}\hline
\begin{tabular}{c}Brane \\Probe \end{tabular} & Theory & Background & 
\begin{tabular}{c} World-Volume \\ Theory \end{tabular} & 
\begin{tabular}{c}Vacuum \\Moduli \\Space \end{tabular} \\ \hline
D5 & Type IIB & $\IR^{1,5} \times CY2$ & (5+1)-d $\CN=1$ Gauge Theory & CY2 \\ \hline
D3 & Type IIB & $\IR^{1,3} \times CY3$ & (3+1)-d $\CN=1$ Gauge Theory & CY3 \\ \hline
M2 & M-theory & $\IR^{1,2} \times CY4$ & (2+1)-d $\CN=2$ Chern-Simons & CY4 \\ \hline
\end{tabular}
\caption{{\sf Brane probes  and associated world-volume physics in various back-grounds.}}
\label{t:embed}
\end{center}
\end{table}

A crucial feature for all the brane embeddings in Table \ref{t:embed} is that in the toric case they are all described by brane tilings. The first case, with CY2, is described by one-dimensional tilings, i.e., brane intervals and thus brane constructions following the work of \cite{Hanany:1996ie}.
The second case is the well established two-dimensional brane tilings which uses dimer techniques to study supersymmetric gauge theories \cite{Hanany:2005ve,Franco:2005rj,Kennaway:2007tq}. The third case is the newly proposed construction \cite{Hanany:2008cd} of Chern-Simon theories.

It is perhaps na\"{\i}vely natural to propose three-dimensional tilings for the case of M2 branes probing CY4 but in fact it turns out not to be as useful as it may initially seem. These three-dimensional tilings have been nicely advocated in the crystal model \cite{crystal}. The main issue perhaps is the current shortcoming of this model to identify the gauge groups with a symplex as it is done for the tilings in dimensions one and two.
In the one-dimensional case for toric CY2, the gauge group is identified with an edge of the tiling, and the matter content, with nodes. For the two-dimensional case for toric CY3, the gauge fields, matter fields, and interactions are respectively identified with faces, edges and nodes of the tiling. But for the proposed crystal model there is no such simple interpretation yet known. 

We are thus led, for now, to keep on the path of two-dimensional tilings, while bearing in mind that the data needed to specify a QCS theory is given by gauge groups, 
matter fields, and interactions, as well as the additional data of the CS levels for the gauge groups. These nicely map, respectively, to tiles, edges and nodes, while the corresponding CS levels are given by fluxes on the tiles.
It would be interesting to check if this correspondence between tilings in one and two dimensions, i.e., for toric Calabi-Yau $n$-folds with $n=2,3,4$, can be extended to possibly higher dimensional tilings and perhaps higher dimensional Calabi-Yau spaces.

The cases for Calabi-Yau two and three-folds are well-established over the past decade. These are affine complex cones over base complex curves and surfaces, or real cones over real, compact, Sasaki-Einstein three and five manifolds. Perhaps the most extensively studied are, inspired by phenomenological concerns, D3-branes and Calabi-Yau three-folds and the widest class studied therein are {\em toric Calabi-Yau cones}. A rather complete picture for both the forward and inverse algorithms, as well as the unifying perspective of brane tilings and dimer models have emerged over the last decade. Ricci-flat metrics have even been found for infinite families within the class of these non-compact spaces.

Another crucial family of Calabi-Yau threefold cones affords a clear construction and the world-volume physics has been intensely investigated (cf.~e.g.~\cite{dps}). The base surfaces here are so-called {\em del Pezzo} surfaces which afford positive curvature, so that the appropriate cones over them have just the right behaviour to make the affine threefold have zero Ricci curvature. More precisely, these surfaces are $dP_n$, which is $\IP^2$ blow-up at $n$ equal to zero up to eight generic points, or the zeroth Hirzebruch surface $F_0 := \IP^1 \times \IP^1$. In fact, the cones over $F_0$ and $dP_{n=0,1,2,3}$ are toric, whereby making these five del Pezzo members of particular interest. The $(3+1)$-dimensional gauge theories for these were first constructed in \cite{Feng:2000mi}, giving rise to such interesting phenomena as toric duality and tilings.

Indeed, all toric gauge theories in $(3+1)$-dimensions obey a remarkable topological formula: take the number of nodes in the quiver, the number of fields and the number of terms in the superpotential; their alternating sum vanishes. This is a key for the powerful brane tiling (dimer model) description  (cf.~review in \cite{Kennaway:2007tq}) of these theories. Interestingly, this relation is still obeyed for the myriad of all known $(2+1)$-dimensional QCS theories to date, and suggests that a planar brane tiling may still be the underlying principle behind theories living on M2-branes probing affine Calabi-Yau fourfolds. The richness of the $(3+1)$-dimensional theories beckon for their analogous and extensions to the $(2+1)$-dimensional case.

It is therefore a natural and important question to ask what are the corresponding geometries for Calabi-Yau four-folds and physically, what are the associated $(2+1)$-dimensional QCS theories on the M2-brane world-volume. That is, what are the (smooth) toric complex three-folds which admit positive curvature?
Based on the ample experience with and the wealth of physics engendered by the aforementioned five del Pezzo cases for three-folds, these four-folds could hold a key toward understanding QCS and M2 theories.

It is the purpose of the current short note, a prologue to \cite{future}, to present the {\it dramatis personae} onto the stage and to introduce some rudiments of their properties as well as to initiate the first constructions of the QCS physics associated thereto. Indeed, complex manifolds admitting positive curvature are in general called {\em Fano varieties} of which the del Pezzo surfaces are merely the two-dimensional examples. We shall see that a complete and convenient classification exists for the smooth toric Fano threefolds over which Calabi-Yau four-fold cones can be established; we shall take advantage of the existing data and use the forward algorithm to explicitly construct the quivers, superpotentials and Chern-Simons levels for some cases. A companion paper, of substantially more length and in-depth analysis \cite{future}, will ensue in the near future. It is our hope that the 18 characters to which we draw your attention shall, in due course, become as familar as the del Pezzo family to the community.

\section{Fano Varieties}
Fano varieties are of obvious importance: these are varieties which admit an ample anti-canonical sheaf; thus, whereas Calabi-Yau varieties are of zero curvature, they are of positive curvature\footnote{
Recently, lower bounds on the Ricci curvature of Fano manifolds have been found \cite{bound}.
}. Therefore, not only could Fano varieties constitute cycles of positive volume that can shrink inside a Calabi-Yau, so too, could they provide local models of Calabi-Yau of a higher dimension. This second case is perhaps of more interest in the brane-probe scenario where the transverse directions to the branes are affine, non-compact Calabi-Yau spaces. In particular, one could construct an affine complex cone over a Fano $n$-fold, so as to construct a Calabi-Yau $(n+1)$-fold and the branes then reside at the tip of the cone. This situation has become well-known to the AdS/CFT correspondence.

What are explicit examples of Fano varieties? In complex dimension one, there is only $\IP^1$, the sphere, which obviously has positive curvature. In dimension two, they are called del Pezzo surfaces. In particular, they are $\IP^2$, as well its blowup $dP_n$ at $n=1$ up to $n=8$ generic points thereon, and the zeroth Hirzebruch surface $\IF_0 := \IP^1 \times \IP^1$. Of these 10, $\IP^2$, $\IF_0$ and $dP_n$ for $n=1,2,3$ admit a toric description. These have been used extensively in constructing gauge theories on the D3-brane world-volume \cite{Feng:2000mi,Feng:2001bn}, the moduli spaces of these theories are correspondingly local Calabi-Yau threefolds. 

We point out that, of course, the afore-mentioned are {\it smooth} Fano varieties. Indeed, we can readily construct affine Calabi-Yau spaces which are singular cones. For example, for complex dimension one, we indeed have the smooth $\IP^1$, leading to the affine Calabi-Yau 2-fold $\IC^2/\IZ_2$, with the corresponding quiver gauge theory in $(5+1)$-dimensions, but we also have any of the famous ADE singularities given by $\IC^2$ quotiented by a discrete subgroup of $SU(2)$ which give rise to well-known gauge theories. In complex dimension 2, we have $\IP^2$, corresponding to the affine Calabi-Yau 3-fold $dP_0 = \IC^3 / \IZ_3$; however, any $\IC^3/\IZ_n$ is just as good with a singular base Fano 2-fold in a weighted projected space. 

Our chief interest lies in the situation of dimension three. These Fano threefolds can give rise to Calabi-Yau four-folds which can then be probed by M2-branes in order to arrive at quiver Chern-Simons (QCS) theories on their world-volume. A classification of the Fano vareities was achieved in the 80's \cite{fano3}; there is a wealth thereof. Our particular interest, will once more be on the toric Fano threefolds where such techniques as tilings and dimers will be conducive. Toric Fano threefolds have been studied in \cite{toricfano3,toricfano4}.
In dimension $n$, an obvious general class of toric Fano $k$-folds is $\prod\limits_{j} \IP^{k_j}$ where $\{k_j\}$ is a partition of $n$, i.e., $n = \sum\limits_j k_j$.

With the rapid advance of computer algebra and algorithmic algebraic geometry, especially in applications to physics (cf.~\cite{alg}), even non-smooth Fano varieties can be classified\footnote{Indeed, in any dimension $d$, it is known there are a finite number of {\it smooth} Fano varieties \cite{smooth}.} \cite{class}. A convenient data-base has been established whereby one could readily search within an online depository\footnote{
We are grateful to Richard Thomas for revealing this treasure trove to us.
} \cite{database}.

\subsection{Smooth Toric Fano Threefolds}\label{s:18}
Given the enormity of the number were we to allow singularities - against which, physically, there need be no prejudice - and inspired by the 2-fold case of the del Pezzo surfaces all being smooth, we shall henceforth restrict our attention to the {\it smooth} toric Fano threefolds. In the parlance of toric geometry, the corresponding cone is called regular. There is a total of 18 such threefolds, a reasonable set indeed. We will adhere to the standard notation of \cite{toricfano4} wherein the family is tabulated and also to the identifier with the database \cite{database} for the sake of canonical reference. This is presented in Table \ref{t:fano3}.

\paragraph{Toric Data: }
Some detailed explanation of the nomenclature in Table \ref{t:fano3} is in order. The toric data is such that the columns are vectors which generate the cone of the variety; in the D-brane context, this has become known as the $G_t$ matrix. Note that each is a 3-vector, signifying that we are dealing with threefolds. Moreover, the point $(0,0,0)$ is always an internal point. This property is equivalent to the Fano condition. Indeed, as we recall from \cite{Feng:2000mi}, the del Pezzo surfaces all have a single internal point. The explicit topology of each space is also given, following \cite{toricfano4}. 

Indeed, our interest in (compact) Fano threefolds is that the complex cone thereupon is a (non-compact) affine Calabi-Yau fourfold which M2-branes may probe. Going form the data in the table to the fourfold is simple, we only need to add one more dimension, say, a row of 1's to each of the matrices. In such cases, the geometry will be cones over what is reported in the third column. In the physics literature there have been several cases which have been studied in considerable depth and detail: the cone over $\IP^3$ is the orbifold $\IC^4 / \IZ_4$, the Sasaki-Einstein 7-fold (a homogeneous space which is a circle fibration over the $\IP^1 \times \IP^2$), which is a {\it real} cone over $\cB_4$ is dubbed $M^{1,1,1}$ (q.v.~\cite{Hanany:2008cd}) and the real Sasaki-Einstein cone over $\cC_3$ is called $Q^{1,1,1} / \IZ_2$ (cf.~e.g.~\cite{Franco,Ami,Franco:2008um}).

\paragraph{Fibrations and Bundles: }
We, of course, recognise $\IP^3$ (succeeding the sequence of $\IP^1$ in dimension 1 and $\IP^2 = dP_0$ in dimension 2) and the natural generalisation $\IP^1 \times \IP^1 \times \IP^1$ of $\IF_0$. Indeed, in $k$ complex dimensions, 
$\IP^k$ and $(\IP^1)^{\times k}$ are always smooth, toric and Fano.
The toric del Pezzo surfaces $dP_{0,1,2,3}$ also appear in the Table, either in direct product or as various fibers. The notation $\IP(~)$ means projectivisation so as to manufacture a compact project threefold. Indeed, we are primarily interested in the {\it affine} Calabi-Yau four-fold cone over these Fano threefolds, so the spaces in which we have interest do not need this projectivisation; we have included them for consistency of notation in that we are discussing the Fano threefolds in this section.

Therefore, the cone in a sense undoes the said projectivisation and the fourfold is simply the total space of the fibration. 
For example, $\cB_1$ is $\IP(\cO_{\IP^2} \oplus \cO_{\IP^2}(2))$; here, $\cO_{\IP^2}(d)$ is a line bundle\footnote{
Of course, in line with standard notation $\cO$ is the structure sheaf, or the line bundle of degree 0.} of degree $d$ over $\IP^2$, hence the fiber of $\cO_{\IP^2} \oplus \cO_{\IP^2}(2)$ is of dimension $1+1=2$, which together with the base $\IP^2$, dictates the total space as being of dimension $2+2=4$. Subsequently, the projectivisation is of dimension $4-1=3$, as required. The actual affine Calabi-Yau fourfold is simply the total space $\cO_{\IP^2} \oplus \cO_{\IP^2}(2)$.

\paragraph{Symmetries: }
One piece of information, obviously of great importance, is the symmetry of the variety, which is encoded in the world-volume physics, either manifestly or as hidden global symmetries \cite{Forcella:2008bb,Feng:2002zw,Franco:2004rt}. Inspecting the toric diagrams, we readily see that our list of Fano threefolds affords the following symmetries. The most symmetric case is, of course, $\IP^3$, the cone over it has a full $U(4)$, acting as unitary transformations on the four co\"ordinates. Next, $\cB_1$ and $\cB_2$ both have $SU(3) \times U(1)^2$, with $SU(3)$ acting on the base $\IP^2$ and $U(1)$ for each fiber. Similarly, $\cB_3$ has symmetry $SU(2)^2 \times U(1)^2$, with $SU(2)$ for the base $\IP^1$, another $U(2)$ for the 2 identical line bundles $\cO_{\IP^1}$ and one more $U(1)$ for $\cO(1)_{\IP^1}$. Likewise, $\cB_4$ has $SU(3) \times SU(2) \times U(1)$, with the $SU(3)$ and $SU(2)$ for the $\IP^2$ and $\IP^1$ respectively and $U(1)$ for the cone which gives the affine Calabi-Yau 4-fold. Proceeding along the same vein, $\cC_1$, $\cC_4$ and $\cC_5$ share the symmetry $SU(2)^2 \times U(1)^2$, $\cC_2$ has $SU(2) \times U(1)^3$ and $\cC_3$ has $SU(2)^3 \times U(1)$. All remaining cases, viz., the $\cD$'s, $\cE$'s and $\cF$'s, are of symmetry $SU(2) \times U(1)^3$.

Note that the rank of the group of symmetries must total to 4 because we are dealing with a toric (affine) Calabi-Yau 4-fold. Indeed, one $U(1)$ factor of the symmetry is the R-symmetry and the remaining rank 3 symmetry, a global mesonic symmetry (cf.~\cite{Forcella:2008bb}) and
there could be possible additional $U(1)$-baryonic symmetries.
We have summarised these mesonic symmetries in the last column of Table \ref{t:fano3}, under the entry $Sym$. Unless explicitly written, we have used the short-hand notation that
\[
[3^{k_3}, 2^{k_2}, 1^{k_1}] := SU(3)^{k_3} \times SU(2)^{k_2} \times U(1)^{k_1}
\ .
\]

We note that the three cases of there being only a {\it single} $U(1)$ symmetry, viz., $\IP^3$ (for which $U(4)$ contains the $U(1)$), $\cB_4$ and $\cC_3$, are products of projective spaces corresponding to the three partitions of 3. The corresponding QCS theories for these have been already constructed in the literature. This is perhaps unsurprising given the high degree of symmetry for these spaces.

\paragraph{Some Geometrical Data: }
We have also listed, to the rightmost of the Table, some geometrical data, such as topological invariants. In particular, we tabulate the second Betti number $b_2$ and the genus $g$. Indeed, $b_2 = E - 3$, where $E$ is the number of external points in the toric diagram, or, since there is alway a single internal point as discussed above, $E$ is the number of columns of $G_t$ minus 1.
Now, recall that in the D3-brane probes on Calabi-Yau threefold case,
the external vertices count the conserved anomaly-free global charges of the $(3+1)$-dimensional gauge theory. Each external vertex in the toric diagram is a divisor and its corresponding charge gives rise to a basis for the set of mesonic and baryonic charges: one of which is the R-symmetry, three of which are mesonic and the remaining $E-4$ charges are baryonic.

However, in our present case of M2-branes probing the Calabi-Yau fourfold, the world-volume Chern-Simons theory in $(2+1)$-dimensions has no notion of anomaly\footnote{
An exception to this is the parity anomaly where one starts with a theory that has no CS terms and one-loop perturbation theory generates a non-zero CS term. Since the CS term is odd under parity, one says that parity is conserved in the classical level but broken by a one loop effect, hence anomalous. This is the only instance in which one can have anomalies in $(2+1)$-dimensions. Nevertheless, all the theories we deal with are protected by supersymmetry and, as long as the ranks are equal, the CS levels do not get quantum corrections (cf.~\cite{Aharony:1997bx}).
} 
and hence there is no distinction between anomalous and anomaly free baryonic charges. Thus, $b_2$ seems to be counting the number of baryonic charges if we extend the analogy from the $(3+1)$-dimensional situation. 

On the other hand, a conserved baryonic charge corresponds to a gauge field in AdS. This is counted by the number of 2-cycles in the Sasaki-Einstien 7-fold (SE7), given by the 3-form on each 2-cycle. The number of 2-cycles in the SE7 is equal to the number of 5-cycles by Poincar\'e duality, which is in turn equal to the number $E$ of external points in the toric diagram subtracted by 4.
That is, the baryonic symmetries also afford a nice geometrical interpretation here: the number of columns of $G_t$ is $E+1$, the number of baryonic symmetries is then $E-4$, signifying $U(1)^{E-4}$ (cf.~section 2 of \cite{Ami}, also \cite{Imamura:2009ph}). Then, since the second Betti number is $E-3$, we have the number of baryonic symmetries as the topological quantity $b_2 - 1$.
\comment{
Therefore, there seems to be a mismatch of 1 between counting with $b_2 = E-3$ and with 3-forms, which is $E-4$.
For example, for $\cC_3 = \IP^1 \times \IP^1 \times \IP^1$ in the table, we have 6 external vertices and there should be 2 baryonic charges while $b_2$ is 3.
This leaves a puzzle as to whether the remaining one corresponds to a baryonic charge or has a different r\^ole. Nevertheless, if we also included the internal vertex, we have 4, which is a good number since it counts the number of mesonic charges, including R-symmetry \cite{Hanany:2008fj}.
Going back to the $(3+1)$-dimensional theory, we recall that each internal vertex in the toric diagram gives rise to 2 anomalous baryonic charges. It is not clear what is the analogous effect in $(2+1)$-dimensions. 
}

Next, let us discuss the genus $g$. Note that a polarisation can be chosen as the ample anti-canonical sheaf $A = K_X^{-1}$, which, due to its ampleness, can be used to embed into a projective space. In turns out that this embedding is of degree $d = c_1(X)^3$ into $\IP^{g+1}$ such that $d = 2g - 2$. Of physical importance is that the $g+2$ homogemeous co\"ordinates of the ambient $\IP^{g+1}$ constitute $g+2$ gauge invariant chiral operators which parametrise the supersymmetric vacuum moduli space, with the relations satisfied amongst them providing the explicit equation thereof. In short, the number of generators of the moduli space is $g+2$.

\paragraph{Hilbert Series: }
Now, it was first pointed out in \cite{Benvenuti:2006qr,Feng:2007ur} that the Hilbert series of an algebraic variety is central toward understanding the gauge invariant operators of the gauge theory living on the branes probing the variety. For our purposes, this is a rational function which is the generating function for counting the spectrum of operators; it could be multi-variate, having a number of ``chemical potentials'', which we call the refined Hilbert series, or it could depend on a single grading, which we call the unrefined Hilbert series. In particular, cones over the Fano two-folds, i.e., the del Pezzo surfaces, have an elegant expression for their unrefined Hilbert series. We recall, from \S3.3.1 of \cite{Benvenuti:2006qr}, that for the $n$-th del Pezzo, of degree $9-n$, it is $f(t; dP_n) = \frac{1+(7-n)t+t^2}{(1-t)^3}$ $(n = 0, \ldots, 8)$; 
Note that $\IF_0$ has the same unrefined Hilbert series as that of $dP_1$; though, the refined, multi-variate, Hilbert series does differentiates the two.

The unrefined Hilbert series, computed for the canonical embedding stated above, is also presented in \cite{database}, though perhaps not of immediate use since they are given as series expansions. We have recomputed these as rational functions. By inspection, a succint equation, similar to the del Pezzo case, exists:
\begin{equation}
f(t;~X) = \frac{1 + (g-2)t + (g-2)t^2 + t^3}{(1-t)^4} = 
\sum_{n=0}^\infty \frac{t^n}{6}(2n+1)((g-1)n^2+(g-1)n+6)
\ , 
\end{equation}
where $g$ is the genus of $X$.

In the special cases where the fano threefold $X$ is the product of $dP_n$ with $\IP^1$, the genus turns out to be $28 - 3n$. Whence, the number of generators of the moduli space is $30 - 3n = 3(10-n)$; the 3 corresponds to the $\IP^1$ factor and the $10-n$ refers to the $dP_n$ factor.

\begin{table}
\[
\hspace{-1cm}
\begin{array}{|c|c|c|c|c|}
\hline
 & \mbox{Id of \cite{database}} & G_t: \mbox{ Toric Data} & \mbox{Geometry}
 & (b_2,g, Sym) \\
\hline \hline
\IP^3 & 4 & 
\tmat{1 & 0 & 0 & -1 & 0 \\
 0 & 1 & 0 & -1 & 0 \\
 0 & 0 & 1 & -1 & 0} & \IP^3 & (1,33,U(4)) 
	\\
	\hline 
\cB_1 & 35 &
\tmat{
 1 & 0 & 1 & -1 & 1 & 0 \\
 0 & 1 & 1 & -1 & 1 & 0 \\
 0 & 0 & 2 & -1 & 1 & 0
} & \IP(\cO_{\IP^2} \oplus \cO_{\IP^2}(2)) & 
 (2,32,[3,1^2]) 
	\\
	\hline
\cB_2 & 36 &
\tmat{
 1 & 0 & 0 & -1 & -1 & 0 \\
 0 & 1 & 0 & -1 & 0 & 0 \\
 0 & 0 & 1 & -1 & 0 & 0
} &  \IP(\cO_{\IP^2} \oplus \cO_{\IP^2}(1)) & 
 (2,29, [3,1^2]) 
	\\
	\hline
\cB_3 & 37 &
\tmat{
1 & 0 & 0 & -1 & -1 & 0 \\
 0 & 1 & 0 & -1 & -1 & 0 \\
 0 & 0 & 1 & -1 & 0 & 0
} & \IP(\cO_{\IP^1} \oplus \cO_{\IP^1} \oplus \cO_{\IP^1}(1)) &
 (2,28, [2^2,1^2]) 
	\\
	\hline
\cB_4 & 24 &
\tmat{ 1 & 0 & 0 & -1 & 0 & 0 \\
 0 & 1 & 0 & -1 & 0 & 0 \\
 0 & 0 & 1 & 0 & -1 & 0 } & \IP^2 \times \IP^1  & 
 (2,28,[3,2,1]) 
	\\
	\hline
\cC_1 & 105 & 
\tmat{
 1 & 0 & 1 & -1 & 0 & 1 & 0 \\
 0 & 1 & 1 & -1 & 0 & 1 & 0 \\
 0 & 0 & 1 & 0 & -1 & 0 & 0 } & 
\IP(\cO_{\IP^1 \times \IP^1} \oplus \cO_{\IP^1 \times \IP^1}(1,1) ) & 
 (3,27,[2^2,1^2]) 
	\\
	\hline
\cC_2 & 136 &
\tmat{
 1 & 0 & 0 & -1 & -1 & -2 & 0 \\
 0 & 1 & 0 & -1 & 0 & -1 & 0 \\
 0 & 0 & 1 & -1 & 0 & -1 & 0} &
\IP( \cO_{dP_1} \oplus \cO_{dP_1}(\ell)), \quad 
\left.\ell^2\right|_{dP_1} = 1 & 
 (3,26,[2,1^3]) 
	\\
	\hline
\cC_3 & 62 &
\tmat{
 1 & 0 & 0 & -1 & 0 & 0 & 0 \\
 0 & 1 & 0 & 0 & -1 & 0 & 0 \\
 0 & 0 & 1 & 0 & 0 & -1 & 0 } & \IP^1 \times \IP^1 \times \IP^1 &
 (3,25,[2^3,1]) 
	\\
	\hline
\cC_4 & 123 &
\tmat{
 1 & 0 & 0 & -1 & 0 & -1 & 0 \\
 0 & 1 & 0 & -1 & 0 & 0 & 0 \\
 0 & 0 & 1 & 0 & -1 & 0 & 0 } & dP_1 \times \IP^1 &
 (3,25,[2^2,1^2]) 
	\\
	\hline
\cC_5 & 68 &
\tmat{ 1 & 0 & 0 & -1 & -1 & 1 & 0 \\
 0 & 1 & 0 & -1 & -1 & 1 & 0 \\
 0 & 0 & 1 & -1 & 0 & 0 & 0} & 
\IP(\cO_{\IP^1 \times \IP^1} \oplus \cO_{\IP^1 \times \IP^1}(1,-1) ) & 
 (3,23,[2^2,1^2]) 
	\\
	\hline
\cD_1 & 131 &
\tmat{
 1 & 0 & 0 & -1 & -1 & -1 & 0 \\
 0 & 1 & 0 & -1 & 0 & -1 & 0 \\
 0 & 0 & 1 & -1 & 0 & 0 & 0} & 
\IP^1\mbox{-blowup of }\cB_2 & 
 (3,26,[2,1^3]) 
	\\
	\hline
\cD_2 & 139 &
\tmat{
 1 & 0 & 0 & -1 & -1 & 0 & 0 \\
 0 & 1 & 0 & -1 & 0 & -1 & 0 \\
 0 & 0 & 1 & -1 & 0 & -1 & 0} & \IP^1\mbox{-blowup of }\cB_4 & 
 (3,24,[2,1^3]) 
	\\
	\hline
\cE_1 & 218 & 
\tmat{1 & 0 & 0 & -1 & -1 & 0 & -1 & 0 \\
 0 & 1 & 0 & -1 & 0 & -1 & -1 & 0 \\
 0 & 0 & 1 & -1 & 0 & 0 & 0 & 0} &
dP_2 \mbox{ bundle over } \IP^1 & 
 (4,24, [2,1^3]) 
	\\
	\hline
\cE_2 & 275 &
\tmat{
 1 & 0 & 0 & -1 & 0 & -1 & -1 & 0 \\
 0 & 1 & 0 & -1 & 0 & 0 & 0 & 0 \\
 0 & 0 & 1 & 0 & -1 & 0 & -1 & 0} &
dP_2 \mbox{ bundle over } \IP^1 & 
 (4,23, [2,1^3]) 
	\\
	\hline
\cE_3 & 266 &
\tmat{
 1 & 0 & 0 & -1 & 0 & -1 & 0 & 0 \\
 0 & 1 & 0 & -1 & 0 & 0 & -1 & 0 \\
 0 & 0 & 1 & 0 & -1 & 0 & 0 & 0} &
dP_2 \times \IP^1 & 
 (4,22,[2,1^3]) 
	\\
	\hline
\cE_4 & 271 &
\tmat{
 1 & 0 & 0 & -1 & -1 & -1 & 1 & 0 \\
 0 & 1 & 0 & -1 & 0 & -1 & 1 & 0 \\
 0 & 0 & 1 & -1 & 0 & 0 & 0 & 0} &
dP_2 \mbox{ bundle over } \IP^1 & 
 (4,21, [2,1^3]) 
	\\
	\hline
\cF_1 & 324 &
\tmat{
1 & 0 & 0 & -1 & 0 & -1 & 0 & 1 & 0 \\
 0 & 1 & 0 & -1 & 0 & 0 & -1 & 1 & 0 \\
 0 & 0 & 1 & 0 & -1 & 0 & 0 & 0 & 0} &
dP_3 \times \IP^1 &
 (5,19, [2,1^3]) 
	\\
	\hline
\cF_2 & 369 & 
\tmat{
 1 & 0 & 0 & -1 & -1 & 0 & -1 & 1 & 0 \\
 0 & 1 & 0 & -1 & 0 & -1 & -1 & 1 & 0 \\
 0 & 0 & 1 & -1 & 0 & 0 & 0 & 0 & 0} & 
dP_3 \mbox{ bundle over } \IP^1 & 
 (5,19, [2,1^3]) 
\\
\hline
\end{array}
\]
\caption{The 18 smooth toric Fano threefolds. For full explanation of notation, q.v.~the second paragraph of \sref{s:18} {\it et sequentes}.}
\label{t:fano3}
\end{table}

\section{Reconstructing the Vacuum Moduli Space}
With a current want of an Inverse Algorithm, with or without the aid of dimer technology, it is difficult to systematically find the requisite quiver Chern-Simons theories whose moduli spaces are Calabi-Yau cones over the Fano threefolds listed above, a question certainly of considerable interest. Nevertheless, because the Forward Algorithm is now well-established \cite{Hanany:2008gx}, one could explicitly check whether a certain ansatz theory indeed gives the correct moduli space. Therefore, with a combination of inspired guesses and systematic computer scans, one could hope to find some theories.

\paragraph{Nomenclature: }
In accordance with the notation of \cite{Hanany:2008fj,Franco:2008um}, and emphasising the intimate relation between the $(3+1)$-dimensional gauge theory and the $(2+1)$-dimensional QCS, we denote the latter as follows: let the superpotential and matter content be that of the D3-brane world-volume theory for the Calabi-Yau threefold $X$, then, we keep the same superpotential and quiver, but impose Chern-Simons levels $\vec{k}$, ordered according to a fixed choice for the nodes, while obeying the constraint \cite{Franco:2008um,Hanany:2008gx}
\begin{equation}\label{kcons}
\sum_i k_i = 0 \ , \qquad GCD(k_i) = 1 \ .
\end{equation}
We subsequently run the forward algorithm, the resulting vacuum moduli space is now a Calabi-Yau four-fold and the QCS theory we will denote as $\widetilde{X}_{\vec{k}}$. Note, of course, that the actual 4-fold may be seemingly quite unrelated to $X$.

Furthermore, as always, we let $X_{ij}^a$ denote the $a$-th bi-fundamental field between nodes $i$ and $j$, and let $\phi_i^a$ signify the $a$-th adjoint field for the $i$-th node.

\subsection{Various Candidates}
\paragraph{$\widetilde{dP_0}_{(1,-2,1)}$ and $\cB_4$: }
The quiver and superpotential can be readily recalled from, e.g., \cite{Feng:2000mi} (cf.~also this theory as a QCS from \cite{Hanany:2008cd}); next, we can assign the Chern-Simons levels as $(1,-2,1)$, which indeed satisfies the constraint \eref{kcons}:
\begin{equation}
\ba{l}
W = \epsilon_{\alpha\beta\gamma} X^{(\alpha)}_{12}
  X^{(\beta)}_{23} X^{(\gamma)}_{31} \\
\mbox{CS-levels } = (1,-2,1)
\ea
\qquad
\ba{c} \epsfxsize = 2.5cm\epsfbox{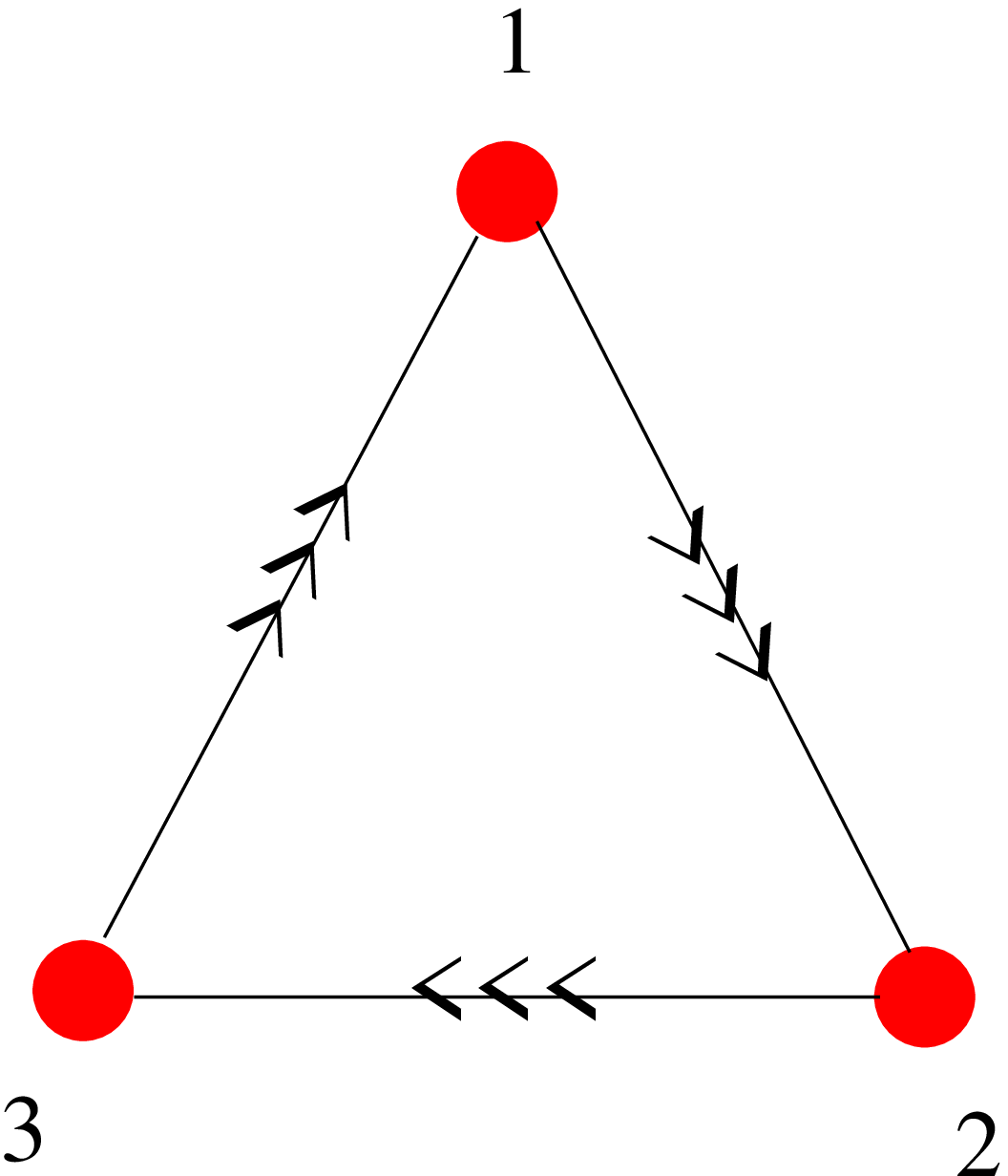} \ea
\end{equation}
Running through the forward algorithm gives us the following charge matrix $Q_t$ and toric diagram $G_t$:
\begin{equation}
Q_t = 
\comment{
{{0, 0, 0, -2, 1, 1}, {-1, -1, -1, 1, 1, 1}}
}
{\scriptsize \left(
\begin{array}{llllll}
 -1 & -1 & -1 & 1 & 1 & 1 \\
 0 & 0 & 0 & -2 & 1 & 1
\end{array}
\right)} \ , \quad
G_t =
\comment{
{{-1, 1, 0, 0, 0, 0}, {0, -1, 1, 0, 0, 0}, {0, 0, 0, 0, -1, 1}, {1, 1,
   1, 1, 1, 1}}}
{\scriptsize \left(
\begin{array}{llllll}
 -1 & 1 & 0 & 0 & 0 & 0 \\
 0 & -1 & 1 & 0 & 0 & 0 \\
 0 & 0 & 0 & 0 & -1 & 1 \\
 1 & 1 & 1 & 1 & 1 & 1
\end{array}
\right)} \ .
\end{equation}

Now, take $\cB_4$, or number 24, of the Fano list from Table \ref{t:fano3}, and consider the affine CY4 cone thereupon, by adding a row of 1's. 
One can readily check, that up to re-ordering of the columns, the two $G_t$ matrices are explicitly related by an $PSL(4; \IZ)$ transformation. This means that the moduli spaces, as affine toric varieties, are isomorphic.

\paragraph{Phases of $F_0$: }
Next, we recall the well-known two phases of the $(3+1)$-dimensional theories for the CY3 over the zeroth Hirzebruch surface:
\begin{equation}
\ba{l}
W_{(F_0)_I} = \epsilon_{ij}\epsilon_{pq}
	{X_{12}}^i{X_{23}}^p{X_{34}}^j{X_{41}}^q ; \\
W_{(F_0)_{II}} = \epsilon_{ij}\epsilon_{mn} X^i_{12}X^m_{23}X_{31}^{jn}- \epsilon_{ij}\epsilon_{mn} X^i_{14}X^m_{43}X_{31}^{jn} \ .
\ea
\qquad
\ba{c} \epsfysize = 3cm\epsfbox{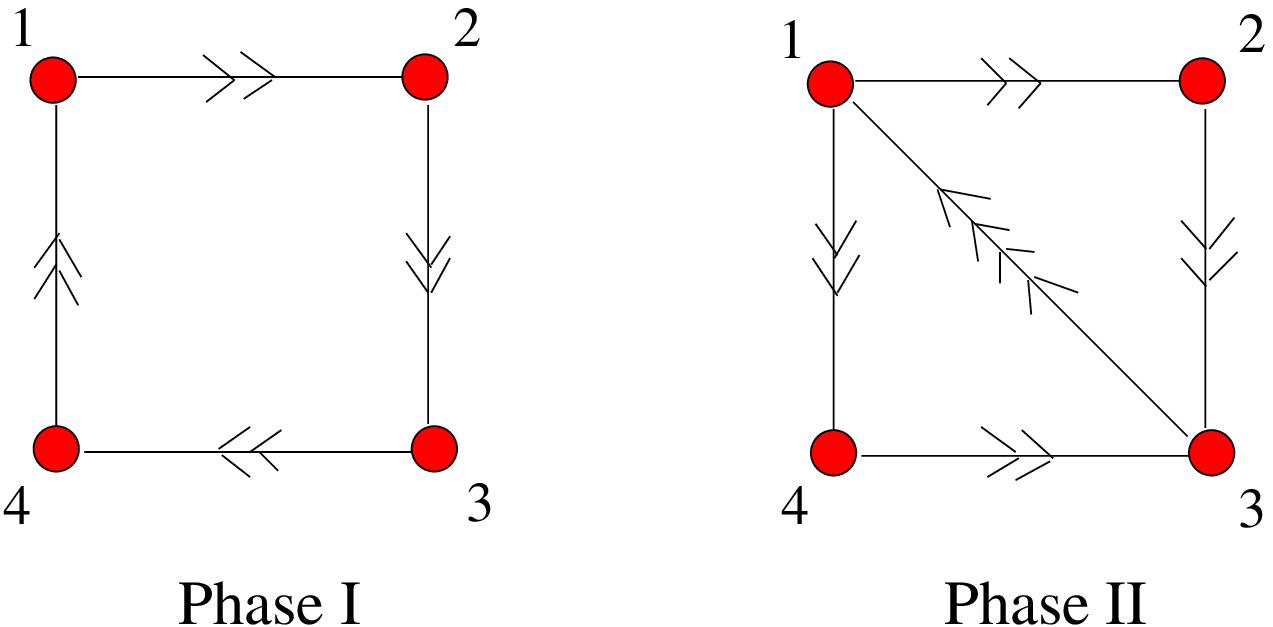} \ea
\end{equation}
There are two toric phases, the first having 8 fields, and the second, 12.

From these progenitors we can obtain quite a few Calabi-Yau fourfold cones, with judicious choices of CS levels. We list these in Table~\ref{t:F0}, running, in each case, the forward algorithm to the theory. The input is the superpotential and quiver of the indicated phase of $F_0$, together with the chosen Chern-Simons levels, and the output, the charge matrix $Q_t$ and toric diagram $G_t$.

\begin{table}[h!]
\[
\hspace{-2cm}
\begin{array}{|c|c|c|c|c|}\hline
F_0 & \mbox{CS Levels } \vec{k} & Q_t & G_t & \sim Cone(X) \\ 
\hline \hline
I & (1,1,-1,-1) & 
{\tiny 
\left(
\begin{array}{llllllll}
 1 & 1 & -1 & 1 & -1 & -1 & 0 & 0 \\
 1 & 1 & 1 & -1 & 0 & 0 & -1 & -1 \\
 0 & 0 & 0 & 0 & -1 & -1 & 1 & 1 \\
 -1 & -1 & 1 & 1 & 0 & 0 & 0 & 0
\end{array}
\right)
}
& 
{\tiny
\left(
\begin{array}{llllllll}
0 & 0 & 0 & 0 & 0 & 0 & -1 & 1 \\
 0 & 0 & 0 & 0 & -1 & 1 & 0 & 0 \\
 -1 & 1 & 0 & 0 & 0 & 0 & 0 & 0 \\
 1 & 1 & 1 & 1 & 1 & 1 & 1 & 1
\end{array}
\right)
}
& \cC_3
\\ \hline
I & (-2,0,1,1) & 
{\tiny
\left(
\comment{
m = Permute[{{-3, 2, -1, -1, 3, 1, 1, -2}, {-2, 2, -2, -2, 0, 3, 
      3, -2}, {0, 1, 0, 0, 0, -1, -1, 1}, {1, 0, -1, -1, 1, 0, 0, 
      0}} // Transpose, {3, 4, 1, 5, 6, 7, 2, 8}] // Transpose
newm = m;
newm[[1]] = 1/2 (m[[1]] - m[[2]] - m[[4]]) + m[[4]];
newm[[2]] = 
 1/2 ((m[[2]] + 1/2 (m[[1]] - m[[2]] - m[[4]]) + m[[4]])/2 - m[[4]] + 
    m[[3]]);
}
\begin{array}{llllllll}
0 & 0 & 0 & 2 & -1 & -1 & 0 & 0 \\
 0 & 0 & -1 & 0 & 0 & 0 & 1 & 0 \\
 0 & 0 & 0 & 0 & -1 & -1 & 1 & 1 \\
 -1 & -1 & 1 & 1 & 0 & 0 & 0 & 0
\end{array}
\right)
}
& 
\comment{
{{0, -1, -1, 0, 1, -1, -1, 1}, {0, 0, 0, 0, -1, 1, 0, 0}, {-1, 1, 0, 
  0, 0, 0, 0, 0}, {1, 1, 1, 1, 1, 1, 1, 1}}
and then add row 1 and 2--> new row 1 
}
{\tiny
\left(
\begin{array}{llllllll}
 0 & -1 & -1 & 0 & 0 & 0 & -1 & 1 \\
 0 & 0 & 0 & 0 & -1 & 1 & 0 & 0 \\
 -1 & 1 & 0 & 0 & 0 & 0 & 0 & 0 \\
 1 & 1 & 1 & 1 & 1 & 1 & 1 & 1
\end{array}
\right)
}
& 
\cC_4
\\ \hline
I & (-2,1,0,1) & 
\comment{
m = {{0, 0, 1, -1, 0, 0, -1, 1}, {-1, -1, -2, 0, 2, 2, -3, 3}, {0, 0, 0, 
  0, -1, -1, 1, 1}, {-1, -1, 1, 1, 0, 0, 0, 0}};
1/2((m[[2]] - m[[1]] + m[[4]])/2 - m[[4]]+ m[[3]]) --> new row 2;
}
{\tiny
\left(
\begin{array}{llllllll}
 0 & 0 & 1 & -1 & 0 & 0 & -1 & 1 \\
 0 & 0 & -1 & 0 & 0 & 0 & 0 & 1 \\
 0 & 0 & 0 & 0 & -1 & -1 & 1 & 1 \\
 -1 & -1 & 1 & 1 & 0 & 0 & 0 & 0
\end{array}
\right)
}
& 
\comment{
gt = {{ -1 , 0 , 0 , -1 , 1 , 0 , 1 , 0 }, {0 , 0 , 0 , 0 , -1 , 1 , 0 , 
  0 }, { -1 , 1 , 0 , 0 , 0 , 0 , 0 , 0}, {1, 1, 1, 1, 1, 1, 1, 1}}
}
{\tiny
\left(
\begin{array}{llllllll}
 -1 & 0 & 0 & -1 & 1 & 0 & 1 & 0 \\
 0 & 0 & 0 & 0 & -1 & 1 & 0 & 0 \\
 -1 & 1 & 0 & 0 & 0 & 0 & 0 & 0\\
 1 & 1 & 1 & 1 & 1 & 1 & 1 & 1
\end{array}
\right)
}
&
\cC_5
\\ \hline
II & (-2,0,1,1) & 
\comment{
m = {{-1, 1, -2, 1, -1, 1, 1, -1, 1}, {-3, 1, 0, 3, -1, 1, 
    1, -1, -1}, {0, 1, 0, 0, 0, -1, -1, 0, 1}, {0, 1, -1, 0, 
    1, -1, -1, 1, 0}, {1, -1, -1, 1, 0, 0, 0, 0, 0}};
m[[2]] = (m[[2]] - m[[1]])/2;
m[[4]] = (m[[4]] - m[[3]]);
m[[3]] = -m[[3]];
m[[1]] = m[[1]] + m[[4]] - m[[5]];
Permute[m // Transpose, {4, 1, 5, 8, 6, 7, 2, 3, 9}] // 
  Transpose // MatrixForm
}
{\tiny
\left(
\begin{array}{lllllllll}
 0 & -2 & 0 & 0 & 1 & 1 & 2 & -2 & 0 \\
 1 & -1 & 0 & 0 & 0 & 0 & 0 & 1 & -1 \\
 0 & 0 & 0 & 0 & 1 & 1 & -1 & 0 & -1 \\
 0 & 0 & 1 & 1 & 0 & 0 & 0 & -1 & -1 \\
 1 & 1 & 0 & 0 & 0 & 0 & -1 & -1 & 0
\end{array}
\right)
}
& 
\comment{
gt = {{0, -1, 1, 0, 0, 0, -1, 0, 1}, {0, 0, -1, 1, 0, 0, 0, 0, 0}, {0, 1, 
  0, -1, -1, 1, 1, 0, -1}, {1, 1, 1, 1, 1, 1, 1, 1, 1}}
and then add 1+2+3 to give new row 3
}
{\tiny
\left(
\begin{array}{lllllllll}
 0 & -1 & 1 & 0 & 0 & 0 & -1 & 0 & 1 \\
 0 & 0 & -1 & 1 & 0 & 0 & 0 & 0 & 0 \\
 0 & 0 & 0 & 0 & -1 & 1 & 0 & 0 & 0 \\
 1 & 1 & 1 & 1 & 1 & 1 & 1 & 1 & 1
\end{array}
\right)
}
&
\cC_4
\\ \hline
II & (-2,1,0,1) & 
\comment{
m = {{0 , 1 , -1 , 0 , 0 , 0 , 0 , 0 , 0 }, {-1 , 2 , 1 , 3 , -1 , 
   0 , 0 , -1 , -3 }, {0 , 1 , 0 , 0 , 0 , -1 , -1 , 0 , 1 }, { 0 , 
   1 , -1 , 0 , 1 , -1 , -1 , 1 , 0}, {1 , -1 , -1 , 1 , 0 , 0 , 0 , 
   0 , 0}};
m[[2]] = (-9 m[[1]] + m[[2]] - m[[3]] + m[[4]] + m[[5]])/4 + m[[1]];
m[[4]] = m[[4]] - m[[3]];
m[[3]] = -m[[3]];
Permute[Transpose[m], {4, 1, 5, 8, 6, 7, 2, 3, 9}] // Transpose
}
{\tiny
\left(
\begin{array}{lllllllll}
0 & 0 & 0 & 0 & 0 & 0 & 1 & -1 & 0 \\
 1 & 0 & 0 & 0 & 0 & 0 & -1 & 1 & -1 \\
 0 & 0 & 0 & 0 & 1 & 1 & -1 & 0 & -1 \\
 0 & 0 & 1 & 1 & 0 & 0 & 0 & -1 & -1 \\
 1 & 1 & 0 & 0 & 0 & 0 & -1 & -1 & 0
\end{array}
\right)
}
& 
\comment{
Permute[{{-1 , 0 , 0 , 1 , 1 , 0 , 1 , 0 , 1}, { 0 , 0 , 0 , 0 , -1 , 0 , 0 , 
  1 , 0}, { 1 , 0 , 0 , -1 , 0 , -1 , 0 , -1 , -1 }, {1 , 1 , 1 , 1 , 
  1 , 1 , 1 , 1 , 1}}//Transpose,{4, 1, 5, 8, 6, 7, 2, 3, 9}] // Transpose
}
{\tiny
\left(
\begin{array}{lllllllll}
1 & -1 & 1 & 0 & 0 & 1 & 0 & 0 & 1 \\
 0 & 0 & -1 & 1 & 0 & 0 & 0 & 0 & 0 \\
 0 & 0 & 0 & 0 & -1 & 1 & 0 & 0 & 0 \\
 1 & 1 & 1 & 1 & 1 & 1 & 1 & 1 & 1
\end{array}
\right)
}
&
\cC_1
\\ \hline
\end{array}
\]
\label{t:F0}
\caption{{\sf The two phases of the $(3+1)$-dimensional gauge theory for the cone over the zeroth Hirzebruch surface $F_0$ beget 4 new QCS theories in $(2+1)$-dimensions, the moduli spaces for which are cones over 4 different Fano 3-folds.}}
\end{table}

In the Table, we have used the notation $\sim Cone(X)$ to mean that it is isomorphic, by an explicit $SL(4; \IZ)$ transformation of the toric diagrams (up to repetition and permutation of the vertices) $G_t$, to the Calabi-Yau fourfold cone over the Fano threefold $X$. Note that the last row of $G_t$ is alway 1, this a consequence of the Calabi-Yau condition. Furthermore, note that the second 2 rows for phase I, corresponding to the F-terms, decouple into diagonal form; this reflects the fact that the master space \cite{Forcella:2008bb} is the direct product of two conifolds. Moreover, the first row of the table, for the theory corresponding to $(\IP^1)^{\times 3}$, has been obtained in \cite{Ami}.

\paragraph{$\widetilde{dP_1}$ and $\cD_1$: }
The theory for the cone over the $dP_1$ surface is again well known. We present it below (note that only two of the three bi-fundamental fields $X_{23}$ group into an $SU(2)$ multiplet and the third is a singlet). Now, if we took the Chern-Simons levels as $(-1,-1,0,2)$, and combining with the standard theory:
\begin{equation}
\ba{l}
W =\epsilon_{ab}X_{13} X_{34}^a X{41}^b +
\epsilon_{ab}X_{42}  X_{23}^a X_{34}^b + 
\epsilon_{ab}X_{34}^3 X_{41}^a X_{12} X_{23}^b \\
\mbox{CS-levels } = (-1,-1,0,2)
\ea
\qquad
\ba{c} \epsfxsize = 2.5cm\epsfbox{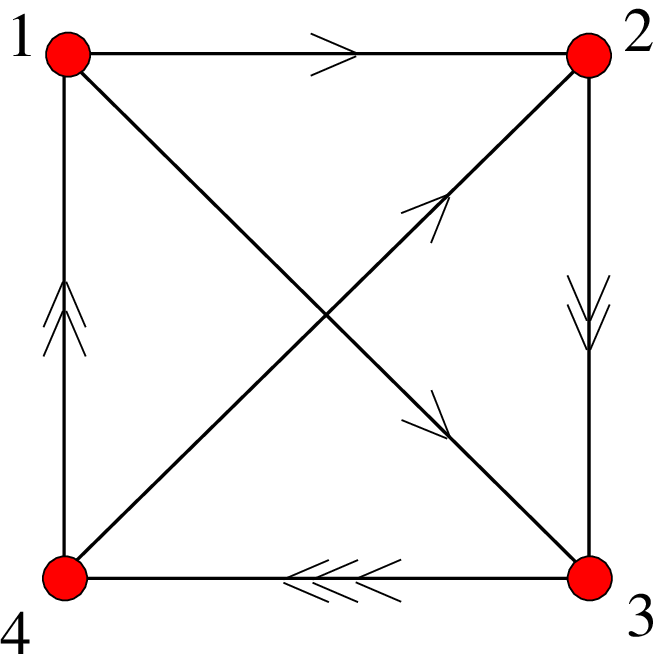} \ea
\end{equation}
then, we find the charge and toric matrices to be:
\begin{equation}
Q_t = {\scriptsize 
\comment{
q = {{-3, -3, 1, 1, 3, 0, 0, 1}, {1, -3, 1, 3, -3, 2, 2, -3}, {-1, 1, 
    1, 0, 0, -1, -1, 1}, {1, 0, -1, -1, 1, 0, 0, 0}};
q = Permute[Transpose[q], {6, 7, 8, 2, 3, 4, 5, 1}] // Transpose;
----intermediate--
(get better row 1 *)
FindInstance[
 And @@ ((-2 <= # <= 2 && Element[#, Integers]) & /@ (a q[[1]] + 
       b q[[2]] + c q[[3]] + d q[[4]])) && a != 0, {a, b, c, d}]
-->
{a -> -1/2, b -> -5/2, c -> -7, d -> -8};
(* then get better row 2*)
FindInstance[
 And @@ ((-2 <= # <= 2 && Element[#, Integers]) & /@ (a q[[1]] + 
       b q[[2]] + c q[[3]] + d q[[4]])) && a != -1/2 && a != 0, {a, b,
   c, d}]
-->
{a -> -3/4, b -> -3/4, c -> -5/2, d -> -2}
------

So: finally
q = {{-3, -3, 1, 1, 3, 0, 0, 1}, {1, -3, 1, 3, -3, 2, 2, -3}, {-1, 1, 
    1, 0, 0, -1, -1, 1}, {1, 0, -1, -1, 1, 0, 0, 0}};
q = Permute[Transpose[q], {6, 7, 8, 2, 3, 4, 5, 1}] // Transpose;
newq = {1/2 (a q[[1]] + b q[[2]] + c q[[3]] + d q[[4]]) /. {a -> -1/2,
      b -> -5/2, c -> -7, 
     d -> -8}, (a q[[1]] + b q[[2]] + c q[[3]] + 
      d q[[4]]) /. {a -> -3/4, b -> -3/4, c -> -5/2, d -> -2}, q[[3]],
    q[[4]]};
newq[[1]] = 
  newq[[1]] + newq[[3]] - (1/3 (newq[[2]] + newq[[3]] - newq[[4]]));
newq[[2]] = 1/3 (newq[[2]] + newq[[3]] - newq[[4]]);
}
\left(
\begin{array}{llllllll}
 0 & 0 & 1 & 1 & 0 & 0 & 0 & -2 \\
 0 & 0 & 0 & 1 & 0 & 0 & -1 & 0 \\
 -1 & -1 & 1 & 1 & 1 & 0 & 0 & -1 \\
 0 & 0 & 0 & 0 & -1 & -1 & 1 & 1
\end{array}
\right)}
\ , \quad
G_t =
\comment{
gt = {{1, 0, -1, 1, 1, 0, 1, 0}, {-1, 0, 0, 0, -1, 1, 0, 0}, {-1, 1, 0, 0, 
  0, 0, 0, 0}, {1, 1, 1, 1, 1, 1, 1, 1}}
}
{\scriptsize 
\left(
\begin{array}{llllllll}
0 & 0 & -1 & 1 & 0 & 1 & 1 & 0 \\
 -1 & 0 & 0 & 0 & -1 & 1 & 0 & 0 \\
 -1 & 1 & 0 & 0 & 0 & 0 & 0 & 0 \\
 1 & 1 & 1 & 1 & 1 & 1 & 1 & 1
\end{array}
\right)
} \ ,
\end{equation}
and resulting moduli space to be $\cD_1$.

\section{Outlook}
In this short note, a prelude to \cite{future}, we have initiated the study of Fano threefolds in the context of M2-branes. In particular, we have presented the classification of all smooth toric Fano threefolds, the cones over which are Calabi-Yau fourfold singularities which the M2-branes could probe. We have computed some preliminary geometrical data, including such quantities as Hilbert series and global symmetries which have recently turned out to be important for the physics of these models.

These 18 spaces are direct analogues of the toric del Pezzo surfaces, which have been the subject of much investigation in the past decade in association with the construction of $(3+1)$-dimensional world-volume quiver gauge theories for D3-branes. It is self-evident that these spaces should be as central to the study of $(2+1)$-dimensional quiver Chern-Simons theories.

For some of these we have identified, using the forward algorithm, the quiver theories whose mesonic moduli spaces are precisely as desired. Such a {\it prima facie} scan has produced 6 as moduli spaces of vacua and they, as with all theories so far produced in the toric M2-brane scenario, obey the planar brane tiling/dimer model condition. It is our hope that systematically all gauge theories for the 18 spaces can be soon geometrically engineered and the corresponding tiling descriptions, prescribed. These and many further details will appear in the companion work of \cite{future}.

\section*{Acknowledgement}
$\ba{l} \epsfxsize = 7cm\epsfbox{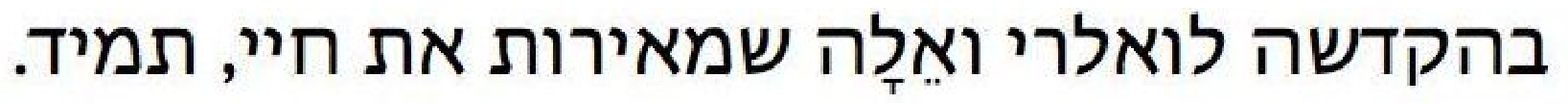} \ea$

{\it Scientiae et Technologiae Concilio Anglicae, et Ricardo Fitzjames, Episcopo Londiniensis, ceterisque omnibus benefactoribus Collegii Mertonensis Oxoniensis, sed super omnes, pro amore Catharinae Sanctae Alexandriae, lacrimarum Mariae semper Virginis, et ad Maiorem Dei Gloriam hoc opusculum Y.-H.~H.~dedicat.}

We are indebted to John Davey, Kentaro Hori, Noppadol Mekareeya, Richard Thomas, Giuseppe Torri, and Alberto Zaffaroni for enlightening discussions. A.H. would like to thank the the kind hospitality, during the initiation of this project, of IPMU in Tokyo, and is further grateful to the University of Richmond, the Perimeter Institute as well as the KITP in Santa Barbara, during the completion. This research was  supported in part by the National Science Foundation
under Grant No. PHY05-51164.


\end{document}